\documentclass[12pt]{article}
\usepackage{cjp}

\begin{document}
% A useful Journal macro
\def\Journal#1#2#3#4{{#1} {\bf #2}, #3 (#4)}

% Some useful journal names
\def\NCA{\em Nuovo Cimento}
\def\NIM{\em Nucl. Instrum. Methods}
\def\NIMA{{\em Nucl. Instrum. Methods} A}
\def\NPB{{\em Nucl. Phys.} B}
\def\PLB{{\em Phys. Lett.}  B}
\def\PRL{\em Phys. Rev. Lett.}
\def\PRD{{\em Phys. Rev.} D}
\def\PRP{{\em Phys. Rept.}}
\def\ZPC{{\em Z. Phys.} C}
\def\JMP{\em J. Math. Phys.}
\def\CMP{\em Commun. Math. Phys.}

% Some other macros used in the sample text
\def\ep{\epsilon}
\def\vep{\varepsilon}
\def\be{\begin{equation}}
\def\ee{\end{equation}}
\def\bea{\begin{eqnarray}}
\def\eea{\end{eqnarray}}
\def\sla{\raise.15ex\hbox{$/$}\kern-.57em}

\title
{\vspace{-5.0cm}
\begin{flushright}{\normalsize RUHN-99-7}\\
\end{flushright}
\vspace*{2.5cm}
Summary Talk at Chiral '99\footnote{Talk at Chiral '99, Sept 13-18, 1999,
Taipei, Taiwan.}
}
\author{H. Neuberger}
\address{Department of Physics and Astronomy\\
	Rutgers University\\
	Piscataway, NJ 08855, USA.}
\maketitle
\begin{abstract}
A brief summary of talks relating to massless lattice fermions is presented.
This summary is not a review and reading it certainly is no substitute to 
reading the various original contributions. 
\end{abstract}
\section{Introduction}
By a rough count this was the third in the Chrial'XX series of conferences
started in Rome in 1992. I guess that a summary ought to first reorder points
made by various speakers by topics and then try to abstract generally
accepted conclusions and identify issues on which agreement is lacking.
As far as the first step, the data was subjected to severe cuts:
there were several very interesting talks outside the narrow topic
of massless fermions on the lattice which I shall not mention. From the
talks that do concern massless lattice fermions I shall pick only
what I think I understood; this is a major cut. I apologize in advance
for omissions and misunderstandings.

The coarsest classification of the topics is into two classes:
\begin{itemize}
\item{Chiral gauge theories.}
\item{Vector-like theories with global chiral symmetries.}
\end{itemize}
\section{Chiral gauge theories}

Let's walk through a list of issues of principle on which I shall present
a status report and, at times, and my personal opinion in a different font.

\begin{itemize}
\item{There exists no complete construction of asymptotically free chiral
gauge theories where the symmetry that is gauged is perturbatively 
non-anomalous.}
\item{There is a disparity in beliefs on whether we have passed the 
point of ``physical plausibility''. By this I mean that, as physicists,
we have established so many features that the remainder of the problem
can be ``shipped over'' to mathematical physics, where in due time 
(hopefully $<\infty$) all hairy technicalities will be nailed down. But,
we no longer have serious doubts about the outcome. Most of us would 
agree, for example, that the RG framework is far beyond physical plausibility.
Nevertheless there is no mathematical proof beyond perturbation theory 
that there always exists a hierarchy of fixed points ordered 
by degrees of stability with appropriate connecting flows, etc.}
\end{itemize}

My opinion is:
\begin{itemize}
\item{\textsl{The older approaches \cite{Leung,Schierholz} still are below 
the point of ``physical plausibility''. On the other hand, 
the new approach is past the point of ``physical
plausibility''. I think many of us here disagree on this assessment.}}
\item{\textsl{There exits only one new approach \cite{neubhere}. 
It is obvious, even if
not represented at this conference, that there are some workers worldwide
that would disagree with this.}}
\item{\textsl{I think that most criticisms 
of the new approach, e.g. \cite{testa},
are rooted in the difficulty to make the new approach look 
completely conventional.}}
\end{itemize}

\subsection{Unconventional features of the new approach}

The new approach is unconventional in that the chiral fermion
determinant is (at the first step at least) not gauge invariant,
but the fermion propagator is gauge covariant. This implies that
the fermion determinant and the fermion propagator are not related
in the conventional manner. In the continuum this issue also exists
although it is hidden behind the overall formal character of the
path integral formulation. Fujikawa, in his work on anomalies associated
this feature with the fermion integration measure rather than
with the determinant but this separation is artificial because we 
see only the product of the ``measure'' and the fermion determinant,
at least to any order in perturbation theory. Nevertheless Fujikawa's
view consists of a deep insight, not as much in the terminology, but
because it tells us precisely what I just mentioned above: the fermion
propagators are well behaved under gauge transformations, only the
fermion determinant is not so (in the anomalous case). 
In diagrams this means that anomalies
only come from triangular fermion loop insertions, and when phrased
in this way it sounds less surprising. But, on the lattice
there is no such thing as an integration measure for fermions:
There are no infinities and Grassmann integration has nothing to do
with measures. So, on the lattice one must do something somewhat
unconventional to get the fermion determinant break gauge invariance
while the fermion propagator does not. In the continuum, 
when anomalies cancel, we can get rid of the gauge violation in
the fermion determinant and we might expect a totally conventional
formulation to hold. There are some conjectures how to ultimately
achieve this on the lattice, but nobody has done it yet. I think
that to actually achieve this in full detail will end up having been
unnecessary.

\begin{itemize}
\item{The new approach requires us to choose bases in subspaces of a finite
(if the lattice is finite) dimensional vector space. This choice depends on 
the gauge background. The definition of the spaces is gauge covariant
but the choice of bases is not. }
\end{itemize}

In my opinion
\begin{itemize}
\item{\textsl{ the ambiguity in phase choice that results from
the above is best interpreted as a descendant 
from an ambiguity in an underlying path integral of
conventional appearance but over an infinite number 
of fermion lattice fields. There is no doubt that this is a possible
interpretation\footnote{Could the Heat Kernel approach of
\cite{ichinose} provide another interpretation ?}, 
because the new construction has been obtained from
a system containing an infinite number of fermions and integrating
all but the lightest out. 
The effective theory governing the lightest fermion
can be formulated directly and then the infinite
number of fermions picture is no longer necessary in the framework
of Euclidean field theory. But, if one wishes to give some argument
for why the theory should be unitary after taking 
the continuum limit and subsequently analytically
continuing to real time, the single known way to date is to go
back to the infinite number of fermion language, where one has 
a familiar form of lattice unitarity, 
at least at a formal level. }}
\end{itemize}

\begin{itemize}
\item{It is at the stage of making the phase choice that the obstructive role
of anomalies shows up. 
It is also at this stage that possibly
new obstructions could come in, ``non-perturbative 
anomalies'' \cite{ZinnJustin}}. 
\end{itemize}

I believe that
\begin{itemize}
\item{\textsl{no such problems will 
occur in many ``good'' theories, but I
won't exclude cases we would deem good today, 
but find out that they are bad
tomorrow. Some complications in finite volume in two
dimensions might contain a hint}} in this direction. 
\end{itemize}

It is important to emphasize that the fermions enter the action bilinearly.
The bilinearity has significant consequences and the entire
new approach is dependent on it. Bilinearity means that all one needs
to know about the fermions is their propagator,
the fermion determinant and the possible 't Hooft vertices, all functions
of the gauge background. In trivial topology, there are no 't Hooft
vertices to worry about, and bilinearity gives a simple prescription
for the result of the integral over fermions for any set of 
fermionic observables. This is the content of Wick's theorem. The 
extension to nontrivial topology with the help of inserting 't Hooft
vertices requires some extra functions (zero modes). If we have the
propagators, the zero modes (when present) and the fermionic
determinant we know all there is to know and whether we also employ
and action and Grassmann integration is a manner of notation but 
not substance. What is unconventional for a lattice theory is that
the fermion propagator does not fully define the fermion determinant.
Just like in Euclidean continuum, it does define the absolute value
of the determinant. The phase of the determinant however
needs to be determined separately. 
The main conceptual obstacle overcome by the overlap
construction was concretely realizing this apparently paradoxical
situation. 

\subsection{Phase choice and fine tuning}

\begin{itemize}
\item{What is missing at the moment in the asymptotically free context 
is a full natural choice of the phase of the chiral determinant making 
it explicit that if anomalies cancel gauge invariance can be exactly 
preserved but, if they are not, such a choice cannot be made by
locally changing some operators.}
\end{itemize}

But, we have some partial results:
\begin{itemize}
\item{If anomalies do not cancel one can show that a good definition
of phase, at least within one framework, is impossible.} 
\item{In the case of $U(1)$, if anomalies do cancel, at least in a rather
formal infinite lattice setting, one can find a good definition of the 
phase of the chiral determinant.}
\end{itemize}

I believe that 
\begin{itemize}
\item{\textsl{
the problem of finding a good phase is almost entirely 
a technical problem. I also believe
that it is a hard technical problem, at least at finite volume.}}
\end{itemize}

Let us now turn to the issue of fine tuning which generated much discussion.
First of all even the concept of fine tuning isn't perfectly well defined.
I'll adopt the following definition: Fine tuning is the need to choose
some functions of field variables which, when viewed as a series in elementary
functions of fields, contain numerical coefficients that have to be of some
exact value, with no deviations admitted. The numerical values of the
coefficients are not directly determined by a symmetry principle.

\begin{itemize}
\item{ The solution to the technical problem of phase choice, according to
all conjectures and results to date, requires fine tuning somewhere.}
\end{itemize}

I believe that
\begin{itemize}
\item{\textsl{if a solution to the technical 
problem exists, that solution
defines a neighborhood, a region in coupling 
space, so that for any point
in it the correct continuum limit will emerge 
after gauge averaging. So, you
only need to be in a good neighborhood, not exactly at its center. This,
in my definition eliminates fine tuning, but we had some disagreements 
both on whether this can work and on whether if it does work it 
really is natural. The basic way this is pictured to work is that
in the anomaly free case one can do a strong coupling type of expansion
in the deviation from the ideal point 
in the center of the neighborhood. One cannot see this work in 
weak coupling perturbation theory.}}
\end{itemize}

\begin{itemize}
\item{Currently there is an effort to define the phase of the chiral
determinant in a perfect way. Kikuakwa's work on the $\eta$-invariant
\cite{Kikukawa}, L{\" u}scher's attempts in the non-abelian 
case, including their respective conjectures are all part of this
effort. The conjectures I presented in my talk are an earlier, somewhat
different attempt in the same direction. In my attempt I tried to restrict
all fine tuning to gauge covariant operators, while in the newer way
one fine tunes at the non-gauge covariant level. }
\end{itemize}

In practice I think one shall need to rely on the existence of the
``good neighborhood'' and try to guess a 
phase choice residing in it. There is
numerical evidence that the Brillouin-Wigner phase convention
(maybe more appropriately termed the Pancharatnam convention),
at least in two dimensions, provides a realistic possibility.

\subsection{Future}
\begin{itemize}
\item{A successful conclusion of any approach to find a perfect
phase choice would constitute a significant result in mathematical
physics.}
\end{itemize}

Some personal opinions:
\begin{itemize}
\item{\textsl{ I am not convinced that 
we need many people working on this.
We should all be happy if this issue 
is taken out of the way by somebody.
The likelihood that new physics would 
emerge from a full solution of 
this problem is not high.}}
\item{\textsl{ Technically, things might simplify if one starts 
by considering more closely a mathematical
construction directly at infinite lattice volume.}}
\end{itemize}

\section{Vector-like gauge theories with massless fermions}

In this area there was a substantial amount of progress recently
and contributions have been both original and coming from many
people. The activity here is closely connected to numerical QCD
and therefore of potential importance to particle phenomenology.

\begin{itemize}
\item{\textsl{ I think in this area there are easier open problems.
On the other hand there are no fundamental open issues even at the
level of mathematical physics (like the phase choice in the chiral
case). We can have confidence in the basic premise that we know
now how to formulate QCD with exactly
massless quarks on the lattice.}}
\end{itemize}

\subsection{Numerical QCD}

We have heard about two basic implementations of the new way
to make fermions massless. 

\begin{itemize}
\item{Domain Wall Fermions, (DWF), the more traditional approach, were
reviewed by Christ \cite{christ}.}
\item{Overlap fermions, a bit newer, were discussed by Edwards,
Liu and McNeile \cite{edw,kfl,mcn}.}
\end{itemize}

What are the advantages of these new methods, when compared to 
employing Wilson fermions, say ?

\begin{itemize}

\item{Small quark masses are attainable without exceptional penalties
and without having to go to staggered fermions with the associated
flavor identification difficulties. But, the price is still high. 
Actually, with DWF we only saw something like ${{m_\pi}\over {m_\rho}}\sim .5$
while we really would like ${{m_\pi}\over {m_\rho}}\sim .25$. To go so low
a prohibitively large number of slices in the extra dimension 
seems to be required \cite{christ}. On the other hand we heard a report
of attaining ${{m_\pi}\over {m_\rho}}\sim .2$ with overlap fermions
\cite{kfl}.}
\item{\textsl{My guess is that the overlap went to lower masses because
of the so called projection technique 
which allows a numerically accurate
representation of the sign function down to very small arguments. This
could be done also with DWF, but would be costly, because the transfer
matrix is more complicated than the Hermitian Wilson Dirac operator. It
would be illuminating if DWF people were to test the projection method
in their framework, only to potentially identify the badness of their implicit 
approximation to the sign function at the origin as a possible source
of the problems they encounter when 
trying to go to lower quark masses.}}
\item{Related to my comment above, we have seen also first steps in
the design of an HMC dynamical simulations method for overlap fermions
incorporating the projection technique \cite{edw}.}
\item{One has very clean lattice versions of topological effects
and the related $U(1)_A$ problem. Both DWF and overlap work give very
nice results. For example, we saw that indeed $U(1)_A$ is not
restored at $T>T_c$ \cite{christ}, that Random Matrix models work as
expected also at non-zero topology \cite{edw} and that the condensate
$\bar\psi \psi$ behaves as expected \cite{christ,edw,kfl}}.
\item{It is potentially very advantageous to have a formulation where
operator mixing is restricted just like in the continuum. This
can provide substantial numerical progress on matrix elements. 
There are good previous results on the Kaon B-parameter and 
surprising new results on ${{\epsilon^\prime}\over{\epsilon}}$
\cite{christ}.}
\item{\textsl{A natural question is then what can be done with the
overlap in this context. There is a big factor difference in the
machine sizes that are applied to DWF versus overlap, so we may have
to wait for quite a while.}}

\item{A cloud on the horizon has been discussed extensively \cite{edw}.
It has to do with the fact that the density of eigenvalues of the
hermitian Wilson Dirac operator $H_W$ at zero seems not to vanish on the 
lattice at any coupling. This might indicate a serious problem since
the definition of the overlap Dirac operator involves the sign function
of $H_W$. The problem also directly affects DWF, making absurdly large
numbers of slices necessary. The overlap permits a simpler fix. 
But, the problem isn't serious so long one works at fixed
physical volume. In that case, taking the scaling law shown by Edwards,
\cite{edw}, we immediately see that, in principle, going with the
lattice $\beta$ to infinity at fixed physical volume will eliminate
the low lying states of $H_W^2$. How to avoid the problem at low values
of $\beta$, say $5.85, 6.0, 6.2$, is an open and practically important
question. Several options were discussed, including changing the pure
gauge action and changing the form of $H_W$. In this context there
might be some relevance in the new exact bounds on the spectrum of
$H_W^2$ which
were not yet complete at the time of the conference. 
These bounds were derived using also eigenvalue flow equations. Such
equations were emphasized by Kerler in his talk \cite{kerler}.}

\item{The main advantage of DWF over overlap fermions is the lower cost
in dynamical simulations. It seems possible to combine the good features
of DWF with those of overlap fermions using various tricks mentioned
by Edwards \cite{edw}. There are many possibilities and we should
be imaginative.}

\end{itemize}

\subsection{Non-QCD}

\begin{itemize}
\item{Kaplan discussed DW formulations of SUSY theories with no matter.
In the continuum, with ${\cal N}=1$ supersymmetry, 
the masslessness of the gaugino
is known to imply supersymmetry at the renormalized level.} 
\item{Going to
higher ${\cal N}$ 
supersymmetries employing dimensional reduction might not work
\cite{kaplan}.}
\item{The fermion pfaffian related to the lattice gluinos was shown
to be non-negative, thus eliminating a potential thorny numerical problem
\cite{kaplan}.}
\item{Lower dimensional theories might provide interesting playgrounds
\cite{kaplan,nagao}. In particular some simple 3 dimensional gauge theories
with massless fermions might have interesting symmetry breaking patterns.}
\end{itemize}

\subsection{Ginsparg-Wilson Relation, Index}

\begin{itemize}
\item{The Ginsparg Wilson relation is an algebraic requirement best
thought of in terms of Kato's pair \cite{neubhere}. We had
some discussion about the GW-overlap equivalence and the role
of the operator $R$ in the GW relation, see 
\cite{tingwair}.}
\item{The following 
formula for the index is reminiscent of the continuum treatment of
Fujikawa. 
\begin{equation}
{\rm Index} = Tr [s f(h^2 )]
\end{equation}
where,
\begin{equation}
h={1\over 2}\left [ {\gamma_5 + {\rm sign} (H_W )}\right ],~~
s= {1\over 2}\left [ {\gamma_5 - {\rm sign} (H_W )}\right ] ,~~ 
h\equiv\gamma_5 D_o,
\end{equation}
and $f(0)=1$. There might be some connection between this and Fujikawa's
talk here \cite{fujikawa}, which centered on the operator $s$ (the formula
$s=\gamma_5 -h = \gamma_5 (1-D_o)$ 
is slightly different because of different conventions involving
factors of two).}
\item{We saw an analytical calculation showing that the lattice reproduces the
correct anomalies even in backgrounds which are non-trivial topologically
\cite{adams}. Previously, this has been checked only numerically and in
two dimensions.}
\end{itemize}

\subsection{Future}

There clearly is more to do and we have some good prospects for progress.
On the numerical front further investigations of ways to implement the
overlap Dirac operator, or of some equivalent object, are called for. While
DWF are easy to visualize, and indeed produce, in the limit
of an infinite number of slices, the sign function of $\log T_W$ where
$T_W$ is a transfer matrix and $\log T_W$ is the same as $H_W$ up to
lattice corrections, I see a danger in the concentration of large amounts
of computer power on this one version of the new way to put fermions
on the lattice. Once too many cycles are invested in DWF, better ways will
get suppressed for a long time and, if any of the hints we are already
seeing develop into serious obstacles, there will be no developed alternatives.
This would cause delays in translating the beautiful theoretical 
progress we are witnessing into better practical number acquisition. 
In short, I urge DWF implementers to be more broad minded; control over a large
machine comes with a large responsibility.

\section{Conclusions}

It is rare that a subfield of theoretical physics 
solves one of its longstanding problems in a direct and ``honest'' 
way, rather than redefining it. Such a rare event has taken place in the
context of lattice fermions. The solution may have implications for physics
beyond the SM, because it is a way to fully regulate a chiral gauge theory,
outside perturbation theory. This lattice theoretical development holds 
promise also for SM phenomenology because it could change substantially 
the methods of numerical QCD. 

At the moment there are some tensions in the field surrounding issues of 
priority and implementation. These problems would get solved if we had:

\begin{itemize}
\item{More imagination.}
\item{More young people.}
\item{More computing power.}
\end{itemize}

\section {Acknowledgments}

My research at Rutgers is partially supported by the DOE under grant
\# DE-FG05-96ER40559. I wish to express my appreciation of the
immense hospitality and great effort invested by the organizers
of Chiral 99 in Taipei. In particular I think I speak
for all of us when I profess the chiral community's indebtedness
to Ting-Wai Chiu for doing so much to produce an inspiring and 
enjoyable meeting.

\end{document}